\documentclass[a4paper,12pt]{article}
\usepackage{amsmath,amssymb}

\begin{document}
\begin{center}
\section*{About the mechanism of matter transfer along cosmic string}

{S.V. Talalov}

\vspace{5 mm}

\small{Department of Theoretical Physics, State University of Tolyatti,\\
 14 Belorusskaya str., Tolyatti, Samara region, 445667 Russia.\\
svtalalov@tltsu.ru}

\end{center}

\begin{abstract}
We consider the quantum capture of nonrelativistic massive particle by the  moving infinite curve (cosmic string in wire approximation).
It is shown that the  cusp  appearing on a string at a certain point due to the string dynamics can make the wave function collapse
 at this point irrespective of the caption place.

{\bf keywords:} {cosmic strings; cuspidal points.}
\end{abstract}

{PACS Nos.: 03.65 Ge;   11.27 +d.}

\vspace{5mm}

 {\bf I.} The role of the cuspidal points \cite{Turok}
  on  cosmic strings \cite{Vil} was explored recently  to  explain the radio bursts in the   Universe \cite{Vach}. 
  In this brief article   we suggest the simple model which demonstrates   matter transfer along the infinite string;  cusps that appear on
 4D string in certain isolated space-time points \cite{KliNik}  will  play the key role here too.
 Let us cosider the infinite non-stationary curve ${\mathbf x}(s,\,t)$ parametrized by parameter $s\in(-\infty,\infty)$ for every time $t$. 
 We suppose that for certain moments $t_1, \dots$  isolated cusps appear on the curve at some points $s_1,\dots$.
 Conditionally, this object can be called as  a  ''cosmic string in wire approximation''  \cite{Ander}.
 For example we may consider the   Nambu - Goto infinite string  in the Minkowski space - time $E_{1,3}$ so that the 
gauge condition $x_0(s,t)\equiv t$ holds (see, for example, \cite{BarNes}). 
For every time $t$ the infinite curve
${\mathbf x}(s) \in E_{3}$ is considered as a source of  potential forces acting on  massive non-relativistic quantum particle. 
Taking into account the possible interpretation in the terms of the  Nambu - Goto  string,
we must explain how we are  going to describe the interaction: initially, the Nambu - Goto string is the relativistic object.  To apply the
the non-relativistic scattering theory, we must do the correct reduction   from Poincar\'e to Galilei group in corresponding 
string model.  This reduction has been made in the work \cite{Tal1} (see also the work \cite{Tal2} for the case with the finite planar string). 
The details of this theory are not important here.

Thus we will  use the non-stationary Schr\"odinger equation to describe the interaction between some particle and the curve ${\mathbf x}(s,\,t)$.
This curve is smooth for all moments $t<\varepsilon$, where $\varepsilon$ is some small positive number.
Regarding the asymtotical behaviour   for the large values of the parameter $s$, we suppose that
\begin{equation}
 \lim_{|s|\to\infty} s^n |{\mathbf x}(s,\,t) - {\mathbf n}_3 s| =0\,, \qquad    \forall\,n=0,1,2,\dots\,.
\label{asymp}
\end{equation}
The vector ${\mathbf n}_3$ is the ort for the  third coordinate axis here. The infinite string with similar boundary conditions 
has been investigated earlier in the article \cite{Tal1}.
The potential is defined by the matrix elements:

 \begin{equation}
\langle \mathbf p \mid \hat V_a(t) \mid \mathbf p' \rangle =
     {\epsilon_a}\, \chi_a(\mathbf p) \chi_a(\mathbf p')
\int\limits^\infty_{-\infty} e^{ -i( \mathbf p - \mathbf p' ) \mathbf x(s,\,t) } g(s)ds\,,
\label{separ_s}
\end{equation}
 where $\chi_a(\mathbf p) = \theta(1/a - |\,\mathbf p|)$ and the constant ${\epsilon_a} <0$ here. 
 The ''form-factor'' $g(s)$ is the arbitrary function from the Swarz space that satisfies following conditions:
  1)$0\le g(s)\le 1$, 2) $g(s)\equiv 1$ $\forall\,s\in [-R,R]$ for some  $R>>1$.
  This function  cuts the interaction for  the domain  $|s| > R$.
 Why has the potential (\ref{separ_s}) been selected?
 Suppose that  parameters $s=s_0$ and $t=t_0$ were fixed. Then the potential 
 $$\langle \mathbf p \mid \hat V_a^0 \mid \mathbf p' \rangle =      {\epsilon_a}\, \chi_a(\mathbf p) \chi_a(\mathbf p')
 e^{ -i( \mathbf p - \mathbf p' ) {\mathbf x}_0  }\,$$
 will be a well-known separable potential for the ''force center'' $ {\mathbf x}_0 = {\mathbf x}(s_0,t_0) $. Moreover the formula
 $$ \lim_{a\to 0} \langle \mathbf r \mid \hat V_a^0 \mid \mathbf r' \rangle 
 \equiv \lim_{a\to 0} {\epsilon_a} f_a(\mathbf r-{\mathbf x}_0)\overline f_a(\mathbf r'-{\mathbf x}_0) =
 \alpha\delta(\mathbf r - {\mathbf x}_0)\delta(\mathbf r - \mathbf r')\,$$
 will be true for the appropriate manner ${\epsilon_a} \to 0$ \quad\cite{BerFad}. Thus the potential (\ref{separ_s})
  will be the potential $V_a^0$ expanded along  the curve $ {\mathbf x} = {\mathbf x}(s)$. 
   Indeed, we consider  the finite parameter $a\sim 0$ and the particles with momentum $|\,\mathbf p| < 1/a$ only. In this case   the non-locality of the separable potential is not essential because the function $f_a(\mathbf r)$ is
  vanishingly small\,\footnote{The function $\chi_a(\mathbf p)$ was selected as the Heviside function $\theta(1/a - |\,\mathbf p|)$ for simplicity only.  We can  redefine    the function $\chi_a(\mathbf p)$ so that the function $f_a(\mathbf r) \equiv 0$ for $r>a$.}
  for all $ r> a$. 
    We avoid the limit $a\to 0$ in this work for the following reasons.
 \begin{enumerate}
 \item We assume that the realistic cosmic strings have finite radius \cite{Vil}. 
 \item On the other hand,  the limit $a\to 0$ leads to the essential (but misplaced here) mathematical  complications.
 Rigorous theory for stationary curve without cuspidal points was developed in the work \cite{Shon}. 
 There is no  theory for the  curve with  cusps.
    \item The potential (\ref{separ_s}) defines correct integral operator in the Hilbert space $L^2(R_3)$ for every time moment $t$.
   This fact   
   allows to explore the non-stationary scattering problem -- the scattering on the moving string.
    \end{enumerate}
 Practically, the separable approximation for the $\delta$-potential has been applied in the
 work \cite{BerFad}, where the  rigorous interpretation of the hamiltonian $-\Delta +\alpha\delta(\mathbf r)$  was  given first.

 {\bf II.}  Let us consider the massive particle   that is infinitely distant from the string  for $t\to -\infty$
 ($m =1/2$, $\hbar = 1$ for subsequent studies).
We suppose that the corresponding state vector $|\psi^-(t)\rangle$  somehow describes  the movement of the particle towards the string. 
Let the state vector  $|\psi(t)\rangle$ describe the state of the considered particle at the finite moment t.
 Than the following integral equation  can be deduced  for the wave function $\psi(\mathbf p,t) =  \langle\mathbf p|\psi(t)\rangle $
 (see \cite{Newt}, for example):
 \begin{equation}
 \label{IntEq1}
 \psi(\mathbf p,t) =  \psi^-(\mathbf p,t)  - 
 i \int\limits_{-\infty}^t dt' \int d^3\mathbf p' e^{-ip^2(t-t')}\langle \mathbf p \mid \hat V_a(t') \mid \mathbf p' \rangle \psi(\mathbf p',t')
 \end{equation}
 
 Our  suppositions will be following:
 \begin{itemize}
  \item the wave function $\psi(\mathbf p,t)$ describes a free particle until the capture happens, the capture  takes place at  some moment $t << 0$;
 \item the string is  the straight line for $t=0$:
  $${\mathbf x}(s,0) \equiv s\,{\mathbf n}_3\,;  $$ 
         \item    $\psi(\mathbf p,0) = \varphi_\varkappa(\mathbf p)$ where the function  $\varphi_\varkappa(\mathbf p)$ will be the 
         solution for the stationary Schr\"odinger equation  with the energy $E= -\varkappa^2$, the potential (\ref{separ_s}) for the straight stationary string and $g(s)\equiv 1$. 
   \end{itemize}

   The function  $\varphi_\varkappa(\mathbf p)$ satisfies the stationary Schr\"odinder equation
 $$-\varkappa^2 \varphi_\varkappa(\mathbf p) = 
 p^2 \varphi_\varkappa(\mathbf p) +\int d^3\mathbf p' \langle \mathbf p \mid \hat V_a(0) \mid \mathbf p' \rangle \varphi_\varkappa(\mathbf p')\,,$$
 the value $ \varkappa = \varkappa(p_3)$ satisfies the equation
 \begin{equation}
 \label{level1}
 1+ 2\pi{\epsilon_a}\int \frac{\chi_a(\mathbf p)dp_1dp_2}{\varkappa^2 + p^{\,2}}\,=0\,.
 \end{equation}
  Finally,
  $$\varphi_\varkappa(\mathbf p) = \frac{\chi_a(\mathbf p)C(p_{\,3})}{\varkappa^2 + p^{\,2}}\,,$$
 where the  function $C(p_{\,3})$ is an arbitrary function that  
 depends on the manner of preparation of the wave packet $\psi^-(\mathbf p,t)$.
  For example, we may select the function  $C(p_{\,3}) $ so that   our particle is located nearly from the space plane $x_3 =0$ for all
 moments  $t<0$. Generally speaking, $\varphi_\varkappa(\mathbf p) \in (L^2(R_3))^\prime$, where the symbol $~^\prime$ denotes the framed Hilbert space.
 
 Thus the following representation  for the wave function $\psi(\mathbf p,t)$ is deduced from the equation (\ref{IntEq1}) and our suppositions 
  ($t\ge 0$):
 
 \begin{eqnarray}
 \label{Psi1}
 \psi(\mathbf p,t) & = & e^{-ip^2t}\varphi_\varkappa(\mathbf p)  -  i \epsilon_a \chi_a(\mathbf p)
 \int\limits_{0}^t dt' e^{-ip^2(t-t')} \int\limits_{-\infty}^\infty ds' g(s') e^{-i \mathbf p\mathbf x (s',\,t')}  I(s',t') \,,\\
 \label{I1}
   I(s,t) & = & \int d^3 \mathbf p \chi_a(\mathbf p) e^{i \mathbf p\mathbf x (s,\,t)}\psi(\mathbf p,t) \,.      
 \end{eqnarray}
For example, in the simplest case  of the rectilinear stationary string   ${\mathbf x}(s,\,t) \equiv  {\mathbf n}_3 s$ 
 ($\forall s, \forall t$) and $g(s)\equiv 1$  the function
 $\psi(\mathbf p,t) =  e^{i\varkappa^2t}\varphi_\varkappa(\mathbf p)$ satisfies the equations (\ref{Psi1}) - (\ref{I1}) 
 identically.

 The function $I(s,t)$
  satisfies the integral equation:
 \begin{equation}
 \label{IntEq2}
 I(s,t) = I_0(s,t) -  
   {\epsilon_a} \int\limits_{0}^t dt'  \int\limits_{-\infty}^\infty ds' g(s') K(s,t;s',t')  I(s',t') \,,
 \end{equation}
 where the absolute term 
 $I_0(s,t) \equiv \int d^3 \mathbf p \chi_a(\mathbf p) e^{i [\mathbf p\mathbf x (s,\,t) - p^2t]}\varphi_\varkappa(\mathbf p)$ and
 the kernel 
 $$ K(s,t;s',t') = i \int d^3 \mathbf p \chi_a(\mathbf p) e^{i[-p^2(t-t') + \mathbf p ( \mathbf x (s,\,t) - \mathbf x (s',\,t') )]}\,.$$

In this brief article we did not set ourselves any investigations of the integral equation (\ref{IntEq2}) as an object. We will use the first Born 
approximation for the wave function  $\psi(\mathbf p,t)$   only; therefore we replace  $I(s,t) \to I_0(s,t)$   in the formula (\ref{Psi1}).

{\bf III.}
As the next step we will discuss the rearrangement of the  wave  function  $\psi(\mathbf p,t)$ for the moment when the 
cuspidal point appeares on the string. Let the space $TE_3$  be the space of the momentum $\mathbf p $.
In accordance with our suppositions the following domain  
${\mathcal Q}\subset TE_3 $ exists for small $\varepsilon_1<\varepsilon$:

$$  {\mathcal Q}:\quad p_1^2+p_2^2 < q(\varepsilon_1) p_3^2 \,, \qquad 
     {\mathbf p}{\mathbf x}^\prime (s,\,t)\not=0\,,\qquad t\in [0,\varepsilon_1]\,, \qquad\forall\,s \,.$$
The function  $q(\varepsilon_1)$ will be a certain continuous function on interval $(0,\varepsilon)$;
because the string is a straight line which coinsides with the  third coordinate axis  for $t=0$, the function 
$q(\varepsilon_1)\to \infty$ for $\varepsilon_1\to 0$.
Therefore for the directions along the string there are no critical points in the   integral (\ref{Psi1}).
Thus for all directions from the domain  ${\mathcal Q}$ the following representation holds:
\begin{equation}
\label{rep1}
\psi(\mathbf p,t) = \psi(\mathbf p,0) + \delta\psi(\mathbf p,t)\,,
\end{equation}
where small variation  $\delta\psi$ will be the  Swarz - like function      
Thus for all moments $t\in [0,\varepsilon]$ 
 the initial location of the particte on the string doesn't change essentially.

Let the  cusp appear on the string at the point $s=s_1$ at the certain moment $t=t_1$.
This fact  means that ${\mathbf p}{\mathbf x}^\prime (s_1,\,t_1)=0$ for all directions so that 
the   integral (\ref{Psi1}) has the critical point (see, for example, \cite{Fed}). The corresponding asympthotics
 of the function $\delta\psi(\mathbf p,t)$ for all $t>t_1$ will be following  (the main term):
\begin{equation}
\label{asimpt1}
 \delta\psi(\mathbf p, t) \sim  const\, \frac{ e^{i{\mathbf p}{\mathbf x}(s_1,\,t)}}{\sqrt{|\mathbf p|}}\,,
\qquad  |\mathbf p|  \to \infty\,.
\end{equation}
The appearance of the cusp means that the function $\delta\psi\in (L^2(R_3))^\prime $ although
$\delta\psi\in L^2(R_3)$ before.
  The asympthotics   (\ref{asimpt1}) leads to the following behaivour of  the 
  Fourier transformation  $\delta\psi(\mathbf x,t)$   near the
  point ${\mathbf x} = {\mathbf x}(s_1,\,t)$ \cite{GelShi}:
  
  $$\delta\psi(\mathbf x, t) \sim  const\,|{\mathbf x} - {\mathbf x}(s_1,\,t)|^{-5/2}\, .$$
  
Thus the cuspidal point  $ = {\mathbf x}(s_1,\,\cdot)$
 arising on the string at some moment  $t=t_1$ leads to the collapse of  the wave function in the neighbourhood of this point.
The detailed  rigorous investigation of this   ''teleportation''  effect  is the interesting problem that demands  separate work.

\end{document}